# Robust State and fault Estimation of Linear Discrete Time Systems with Unknown Disturbances

Bessaoudi Talel[1], Ben Hmida Fayçal[2]
Electrical Engineering Department of ESSTT, Research Unit C3S, Tunis University,
5 Avenue Taha Hussein, BP 56, 1008 Tunis, Tunisia
[1]bessaouditalel@yahoo.fr
[2]faycal.benhmida@esstt.rnu.tn

*Abstract*— **This paper presents a new robust fault and state estimation based on recursive least square filter for linear stochastic systems with unknown disturbances. The novel elements of the algorithm are : a simple, easily implementable, square root method which is shown to solve the numerical problems affecting the unknown input filter algorithm and related information filter and smoothing algorithms; an iterative framework, where information and covariance filters and smoothing are sequentially run in order to estimate the state and fault. This method provides a direct estimate of the state and fault in a single block with a simple formulation. A numerical example is given in order to illustrate the performance of the proposed filter.**

## I. Introduction

In the past few years, the problem of filtering in the presence of unknown inputs has attracted big attention, due to its applications in environment. The unknown input filtering problem has been treated in the literature by different approaches. The first approach assumes that the model for dynamical evolution of the unknown inputs is available. When the properties of the unknown input are known, the augmented state Kalman filter (ASKF) is a solution. To reduce computation costs of the ASKF [1] proposed the two stage Kalman filter where the estimation of the state and unknown input are decoupled. The second approach treats the case when we not have a prior knowledge about the dynamical evolution for unknown input. Kitanidis [2] was the first to solve the problem using the linear unbiased minimum-variance. An extend Kitanidis filter using a paramaterizing technique to obtain an optimal filter (OEF) have been proposed by Darouach et al [3]. Hseih [4] has been developed a robust-two stage Kalman filter (RTSKF) equivalent to Kitanidis filter. An (OMVF) reported by C.S Hsieh [5] have been used in order to developed an optimal minimum variance filter (OMVF) to solve degradation problem encountered in (OEF). Gillijns and De Moor [6] has treated the problem of estimating the state in the presence of unknown inputs which affect the systems model. They have been developed a recursive filter which is optimal in the sense of minimum-variance. This filter has been extended by the same authors for joint input and state estimation to linear discrete-time systems with direct feedthrough where the state and the unknown input estimation are interconnected. This filter is called recursive three step filter (RTSF) [7] and is limited to direct feedthrough matrix with full rank. Cheng et al, [8] proposed a recursive optimal filter with global optimality in the sense of unbiased minimum-variance. This filter is limited to estimate the state. The case of an arbitrary rank has been proposed by Hsieh in [9] the designed optimal filter Known as ERTSF (Extend RTSF). Recently, another technique using a least square method have been proposed by Talel et al, [10] to estimate the state and unknown input.

The Fault Detection and Isolation (FDI) problem for linear systems with unknown disturbances is generally studied, see e.g. Nikoukhah [11], Keller, [12], Chen and Patton, [13,14], Ben Hmida et al, [15]. According to [11], a robust fault detection and isolation in continous-time is developed using the error innovation technique to generate an unbiased white residual signals. The fault is diagnosed by a statistical testing. A new method is developed in [16] to detect and isolate multiple faults appearing simultaneously or sequentially in linear time-invariant stochastic discrete-time systems with unknown inputs [12]. Their methods consist of generating directional residuals using an isolation filter. In [13] the optimal filtering and robust fault diagnosis problem has been studied for stochastic systems with unknown disturbances. The output estimation error with disturbance decoupling is used as a residual signal. After that, a statistical testing procedure is applies to examine the residual and to diagnose faults. Netherless, the simultaneous actuator and sensor fault and state problem is not treated in [13, 14]. Recently, [17] present a new optimal recursive filter for state and fault estimation of linear stochastic systems with unknown disturbances. This method is based on the assumption that no prior knowledge about the dynamical evolution of the unknown disturbances is available. The filter has two advantages: it considers an arbitrary direct feedthrough matrix of the fault and it permits a multiple faults estimations.

In order the overcome this problem, we introduce the square root approach into recursive least square filter. We assume that the unknown disturbances affect only the state equation, while, the fault affects both the state and the output equations where the direct feedthrough matrix has an arbitrary rank and under the specific condition that the process and measurement noise are correlated.

This paper is organized as follows. In Section 2, the problem of fault detection is specifically stated for stochastic system. In section 3, we develop a robust filter. Then, the performance of the designed filter is demonstrated through a simulated example in section 4, followed by few concluding remarks in section 5.

## II. PROBLEM FORMULATION

Consider the linear stochastic discrete-time system in the following form:

$$x_{k+1} = A_k x_k + B_k u_k + F_k^x f_k + G_k d_k + w_k \quad (1)$$

$$y_k = C_k x_k + F_k^y f_k + v_k \quad (2)$$

Where $x_k \in \Re^n$ is the state vector, $y_k \in \Re^m$ is the measurement vector, $u_k \in \Re^r$ is the known input, $f_k \in \Re^p$ is the additive fault vector and $d_k \in \Re^q$ is the unknown disturbances vector. The process noise $w_k$ and the measurement noise $v_k$ are correlated white noise sequences of zero-mean with joint covariance matrix

$$\varepsilon \left[ \begin{bmatrix} v_k \\ w_k \end{bmatrix} \begin{bmatrix} v_k^T & w_k^T \end{bmatrix} \right] = \begin{bmatrix} R_k & S_k^T \\ S_k & Q_k \end{bmatrix} \delta_k \geq 0 \quad (3)$$

With $R_k > 0$, where $\delta_k$ is the unit pulse. The matrices $A_k$, $B_k$, $F_k^x, G_k$, $C_k$ and $F_k^y$ are known and have appropriate dimensions. We assume that $(C_k, A_k)$ is observable, $m \geq p+q$ and the initial state is uncorrelated with the white noises $w_k$ and $v_k$. The initial state $x_0$ is a gaussian random variable with $\varepsilon[x_0] = \hat{x}_0$ and $\varepsilon\left[(x_0 - \hat{x}_0)(x_0 - \hat{x}_0)^T\right] = P_0$ where $\varepsilon[.]$ denotes the expectation operator. The aim of this paper is to design an unbiased minimum-variance linear estimator of the state $x_k$ and fault $f_k$ without any information concerning the fault.

First we represent the process and measurement noise in the following form:

$$\begin{bmatrix} v_k \\ w_k \end{bmatrix} = \underbrace{\begin{bmatrix} R_k^{1/2} & 0 \\ X_k & Q_{x,k}^{1/2} \end{bmatrix}}_{\Upsilon_k} \begin{bmatrix} \tilde{v}_k \\ \tilde{w}_k \end{bmatrix} \text{ with } \begin{bmatrix} \tilde{v}_k \\ \tilde{w}_k \end{bmatrix} \sim (0, I_{m+n}) \quad (4)$$

The matrices $X_k$ and $Q_{x,k}$ satisfy

$$X_k = S_k R_k^{-T/2}, Q_{x,k} = Q_k - S_k R_k^{-1} S_k^T \quad (5)$$

To represent the state information an equation format, we introduce an auxiliary random variable $\tilde{x}_{k/k-1}$ with mean zero and covariance matrix $I_n$, that is $\tilde{x}_{k/k-1} \sim (0, I_n)$. Since we assumed the covariance matrix $P_{k/k-1}$ to be semi-positive definite, we can compute its square root $P_{k/k-1}^{1/2}$ such that:

$$P_{k/k-1} = P_{k/k-1}^{1/2} P_{k/k-1}^{T/2} \quad (6)$$

This square root can be chosen to be upper or lower triangular With $\tilde{x}_{k/k-1}$, $\hat{x}_{k/k-1}$, $d_{k-1}$ and $P_{k/k-1}^{1/2}$ so defined, the variable $x_k$ can be modeled through the following matrix equation:

$$x_k = \hat{x}_{k/k-1} - G_{k-1} d_{k-1} - P_{k/k-1}^{1/2} \tilde{x}_{k/k-1} \quad (7)$$

This equation is called a generalized covariance representation

$$\hat{x}_{k/k-1} = x_k + P_{k/k-1}^{1/2} \tilde{x}_{k/k-1} + G_{k-1} d_{k-1} \quad (8)$$

From equation (1), (2) and (8), we obtain the following set of constraint equations on the unknown $x_k$, $f_k$ and $x_{k+1}$:

$$\begin{bmatrix} \hat{x}_{k/k-1} \\ y_k \\ -B_k u_k \end{bmatrix} = \begin{bmatrix} I_n & 0 & 0 \\ C_k & F_k^y & 0 \\ A_k & F_k^x & -I_n \end{bmatrix} \begin{bmatrix} x_k \\ f_k \\ x_{k+1} \end{bmatrix} + \begin{bmatrix} G_{k-1} & 0 \\ 0 & 0 \\ 0 & G_k \end{bmatrix} \begin{bmatrix} d_{k-1} \\ d_k \end{bmatrix} + \begin{bmatrix} P_{k/k-1}^{1/2} & 0 & 0 \\ 0 & R_k^{1/2} & 0 \\ 0 & X_k & Q_{x,k}^{1/2} \end{bmatrix} \begin{bmatrix} \tilde{x}_{k/k-1} \\ \tilde{v}_k \\ \tilde{w}_k \end{bmatrix} \quad (9)$$

Let this set of equations be denoted compactly by

$$\bar{y}_k = \bar{F}_k \bar{x}_k + \bar{G}_k \bar{d}_k + \bar{L}_k \bar{\mu}_k \quad (10)$$

The weighted least-square problem for the derivation of the square-root filter algorithm by:

$$\min_{\bar{x}_k} \bar{\mu}_k^T \bar{\mu}_k \text{ subject to } \bar{y}_k = \bar{F}_k \bar{x}_k + \bar{G}_k \bar{d}_k + \bar{L}_k \bar{\mu}_k$$

The goal of the analysis of the weighted least-square problem is the derivate of square root solution for the filtred and one step ahead predicted state and fault estimation. Therefore, we will address the numerical transformation involved in solving (10) in two consecutive parts. We start with the derivation of the square-root algorithm for computing the filtered state and fault estimation in section 3.1 and the derivation for the computation of the one-step ahead prediction is presented in section 3.2.

We define the following transformation $T_l^m = \begin{bmatrix} C_k & -I_m & 0 \\ I_n & 0 & 0 \\ 0 & 0 & I_n \end{bmatrix}$,

then using $T_l^m$, resulting transformed set of constraint equation is:

$$T_l^m \bar{y}_k = T_l^m \bar{F}_k \bar{x}_k + T_l^m \bar{G}_k \bar{d}_k + T_l^m \bar{L}_k \bar{\mu}_k$$

Then we have:

$$\begin{bmatrix} C_k \hat{x}_{k/k-1} - y_k \\ \hat{x}_{k/k-1} \\ -B_k u_k \end{bmatrix} = \begin{bmatrix} 0 & -F_k^y & 0 \\ I_n & 0 & 0 \\ A_k & F_k^x & -I_n \end{bmatrix} \begin{bmatrix} x_k \\ f_k \\ x_{k+1} \end{bmatrix} + \begin{bmatrix} C_k G_{k-1} & 0 \\ G_{k-1} & 0 \\ 0 & G_k \end{bmatrix} \begin{bmatrix} d_{k-1} \\ d_k \end{bmatrix} + \begin{bmatrix} C_k P_{k/k-1}^{1/2} & -R_k^{1/2} & 0 \\ P_{k/k-1}^{1/2} & 0 & 0 \\ 0 & X_k & Q_{x,k}^{1/2} \end{bmatrix} \begin{bmatrix} \tilde{x}_{k/k-1} \\ \tilde{v}_k \\ \tilde{w}_k \end{bmatrix} \quad (11)$$

So from (11) we formulate the problem (LS) as follows:

$$\min_{f_k, \hat{x}_k, \hat{x}_{k+1}} \left\| \begin{bmatrix} C_k \hat{x}_{k/k-1} - y_k \\ \hat{x}_{k/k-1} \\ -B_k u_k \end{bmatrix} - \begin{bmatrix} 0 & -F_k^y & 0 \\ I & 0 & 0 \\ A_k & F_k^x & -I_n \end{bmatrix} \begin{bmatrix} x_k \\ f_k \\ x_{k+1} \end{bmatrix} \right\|_{\bar{W}_k}^2 \quad (12)$$

where $\bar{W}_k$ is the weighting matrix chosen as follows:

$$\bar{W}_k = \left( \begin{bmatrix} C_k P_{k/k-1}^{1/2} & -R_k^{1/2} & 0 \\ P_{k/k-1}^{1/2} & 0 & 0 \\ 0 & X_k & Q_{x,k}^{1/2} \end{bmatrix} \begin{bmatrix} C_k P_{k/k-1}^{1/2} & -R_k^{1/2} & 0 \\ P_{k/k-1}^{1/2} & 0 & 0 \\ 0 & X_k & Q_{x,k}^{1/2} \end{bmatrix}^T \right) \quad (13)$$

In the next section we propose to design an unbiased minimum variance linear estimator of the state $x_k$ and the fault $f_k$ without any information concerning the fault $f_k$.

## III. FILTER DESIGN

To solve the problem (12), we propose to decompose it into two parts: a first part to estimate an unbiased minimum variance of the state and fault and a second part to the time update of the filter.

*A Measurement update*

The measurement update is derived from (12) by extracting the rows that depend only on $x_k$ and $f_k$. This yield,

$$\min_{\hat{f}_k, \hat{x}_k} \left\| \begin{bmatrix} y_k - C_k \hat{x}_{k/k-1} \\ \hat{x}_{k/k-1} \end{bmatrix} - \begin{bmatrix} 0 & F_k^y \\ I & 0 \end{bmatrix} \begin{bmatrix} x_k \\ f_k \end{bmatrix} \right\|^2_{\bar{W}_{1,k}} \quad (14)$$

Where $\bar{W}_{1,k}$ denotes the weighting which we give a stochastic interpretation by choosing

$$\bar{W}_{1,k} = \left( \begin{bmatrix} C_k P_{k/k-1}^{1/2} & -R_k^{1/2} \\ P_{k/k-1}^{1/2} & 0 \end{bmatrix} \begin{bmatrix} C_k P_{k/k-1}^{1/2} & -R_k^{1/2} \\ P_{k/k-1}^{1/2} & 0 \end{bmatrix}^T \right) \quad (15)$$

The problem is to determine a linear estimate $\hat{f}_k$, $\hat{x}_k$ of on the given data $y_k$ and $\hat{x}_{k/k-1}$ which have the following form

$$\begin{bmatrix} \hat{f}_k \\ \hat{x}_k \end{bmatrix} = M_k \begin{bmatrix} y_k - C_k \hat{x}_{k/k-1} \\ \hat{x}_{k/k-1} \end{bmatrix} \quad (16)$$

With $M_k \in \Re^{2n \times (m+2n)}$, such that both estimates are a minimum-variance unbiased estimate that is estimate with the properties:

$$\varepsilon(\tilde{x}_k) = \varepsilon(x_k - \hat{x}_k) = 0, \quad (17)$$

$$\varepsilon(\tilde{f}_k) = \varepsilon(f_k - \hat{f}_k) = 0, \quad (18)$$

and the expression below are minimal:

$$\varepsilon\left[(x_k - \hat{x}_k)(x_k - \hat{x}_k)^T\right], \quad (19)$$

$$\varepsilon\left[(f_k - \hat{f}_k)(f_k - \hat{f}_k)^T\right], \quad (20)$$

### a. Unbiased estimation

To obtain an unbiased estimation of state and fault, the matrix $M_k$ must satisfy the following two algebraic constraints:

$$M_k \begin{bmatrix} 0 & -F_k^y \\ I_n & 0 \end{bmatrix} = \begin{bmatrix} 0 & I_n \\ I_n & 0 \end{bmatrix} \quad (21)$$

$$M_k \begin{bmatrix} -C_k G_{k-1} \\ G_{k-1} \end{bmatrix} = \begin{bmatrix} 0 \\ 0 \end{bmatrix} \quad (22)$$

We partitions the matrix $M_k = \begin{bmatrix} M_k^{11} & M_k^{12} \\ M_k^{21} & M_k^{22} \end{bmatrix}$ in the constraint (21) as follows:

$$\begin{bmatrix} M_k^{12} & M_k^{11} F_k^y \\ M_k^{22} & M_k^{21} F_k^y \end{bmatrix} = \begin{bmatrix} 0 & I_n \\ I_n & 0 \end{bmatrix} \quad (23)$$

Hence $M_k^{12} = 0$, $M_k^{11} F_k^y = I_n$, $M_k^{22} = I_n$, $M_k^{21} F_k^y = 0$ (24)

with $M_k^{11} \in \Re^{p \times m}$, $M_k^{21} \in \Re^{n \times m}$. On substituting the constraint equation (22) it can be given as follows

$$\begin{bmatrix} M_k^{11} & M_k^{12} \\ M_k^{21} & M_k^{22} \end{bmatrix} \begin{bmatrix} -C_k G_{k-1} \\ G_{k-1} \end{bmatrix} = \begin{bmatrix} 0 \\ 0 \end{bmatrix} \quad (25)$$

$$\begin{cases} -M_k^{11} C_k G_{k-1} + M_k^{12} G_{k-1} = 0 \\ -M_k^{21} C_k G_{k-1} + M_k^{22} G_{k-1} = 0 \end{cases} \Rightarrow \begin{cases} M_k^{11} C_k G_{k-1} = 0 \\ M_k^{21} C_k G_{k-1} = G_{k-1} \end{cases} \quad (26)$$

The estimators $\hat{f}_k$ and $\hat{x}_k$ are unbiased if $M_k^{11}$ and $M_k^{21}$ satisfy the following constraints:

$$M_k^{11} E_k = T_k \quad (27)$$

$$M_k^{21} E_k = \Gamma_k \quad (28)$$

where

$$E_k = \begin{bmatrix} F_k^y & C_k G_{k-1} \end{bmatrix}, T_k = \begin{bmatrix} I_p & 0 \end{bmatrix} \text{ and } \Gamma_k = \begin{bmatrix} 0 & G_{k-1} \end{bmatrix} \quad (29)$$

The innovation error $\tilde{y}_k$ has the following form

$$\tilde{y}_k = y_k - C_k \hat{x}_{k/k-1} = F_k^y f_k + C_k G_{k-1} d_{k-1} + e_k \quad (30)$$

where

$$e_k = C_k \tilde{\bar{x}}_{k/k-1} + v_k \quad (31)$$

$$\tilde{\bar{x}}_{k/k-1} = A_{k-1} \tilde{x}_{k-1} + F_{k-1}^x \tilde{f}_{k-1} + w_{k-1} \quad (32)$$

**Lemma:** Let rank $(F_k^y) = p$; the necessary and sufficient conditions so that the estimator $\hat{x}_k$ and $\hat{f}_k$ are unbiased as matrix $E_k$ is full colum rank, that is,

$$rank(E_K) = rank\begin{pmatrix} F_k^y & C_k G_{k-1} \end{pmatrix} = p + q.$$

In the next subsection, we propose to determine the gain $M_k^{11}$ and $M_k^{12}$ by satisfying the unbiasedness constraint (17) and (18).

### b. fault estimation

Equation (30) will be written as

$$\tilde{y}_k = E_k \begin{bmatrix} f_k \\ d_{k-1} \end{bmatrix} + e_k \quad (33)$$

Sine $e_k$ not have unit variance and $\tilde{y}_k$ does not satisfy the assumption of the Gauss-Markov theorem [17], the least square solution do not have a minimum-varianve. Netherless, the covariance matrix of $e_k$ has the following form

$$H_k = \varepsilon\left[e_k e_k^T\right] = C_k \bar{P}_{k/k-1}^x C_k^T + R_k, \quad (34)$$

where $\bar{P}_{k/k-1}^x = \varepsilon\left(\tilde{\bar{x}}_{k/k-1} \tilde{\bar{x}}_{k/k-1}^T\right)$

For that, $\hat{f}_k$ can be obtained by a weighted least square (WLS) estimation with a weighting matrix $H_k^{-1}$.

**Theorem :** Let $\tilde{\bar{x}}_{k/k-1}$ be unbiased; the matrix $H_k$ is positive definite and the matrix $E_k$ on is full column rank, then to have a UMV fault estimation, the matrix gain $M_k^{11}$ is given by

$$\left(M_k^{11}\right)^* = T_k E_k^* \text{ where } E_k^* = \left(E_k^T H_k^{-1} E_k\right) E_k^T H_k^{-1} \quad (35)$$

**Proof**: Under that $H_k$ is positive definite and an invertible matrix $L_k \in \Re^{m \times m}$ verifies $L_k L_k^T = H_k$, so we can rewrite (30) as follows:

$$L_k^{-1} \tilde{y}_k = L_k^{-1} E_k \begin{bmatrix} f_k \\ d_{k-1} \end{bmatrix} + L_k^{-1} e_k \quad (36)$$

If the matrix $E_k$ is full column rank, that is, $rank(E_k) = p + q$, then the matrix $\left(E_k^T H_k^{-1} E_k\right)$ is invertible. Solving (36) by an LS estimation is equivalent to solve (33) by WLS solution:

$$\hat{f}_k^* = T_k \left(E_k^T H_k^{-1} E_k\right)^{-1} E_k^T H_k^{-1} \tilde{y}_k \quad (37)$$

suppose that
$$\left(M_k^{11}\right)^* = T_k\left(E_k^T H_k^{-1} E_k\right) E_k^T H_k^{-1} \quad (38)$$

In this way, we consider that $L_k^{-1} e_k$ has a unit variance and (36) can satisfy the assumption of the Gauss-Markov theorem. Hence, (37) is the UMV estimate of $f_k$.

The fault estimation error is given by:
$$\tilde{f}_k = f_k - \hat{f}_k = \left(I - M_k^{11} F_k^y\right) f_k - M_k^{11} C_k G_{k-1} d - M_k^{11} e_k \quad (39)$$

Then, the fault error estimation is rewritten as follows:
$$\tilde{f}_k^* = -\left(M_k^{11}\right)^* e_k \quad (40)$$

from equation (40) we can calculate $\tilde{f}_k$:

Using (34), the covariance $P_k^f$ matrix is given by
$$P_k^{f*} = \varepsilon\left(\tilde{f}_k^* \tilde{f}_k^{*T}\right) = \left(M_k^{11}\right)^* H_k \left(M_k^{11}\right)^{*T} = T_k\left(E_k^T H_k^{-1} E_k\right)^{-1} T_k^T \quad (41)$$

### c. state estimation

In this part, we propose to obtain to obtain an unbiased minimum variance state estimator to calculate the gain matrix $M_k^{21}$ wich will minimise the trace of covariance matrix $P_k^x$ under the unbiasedness constraint (28).

**Théoreme:** Let $E_k^T H_k^{-1} E_k$ be nonsingular, then the state gain matrix $M_k^{21}$ by
$$\left(M_k^{21}\right)^* = \bar{P}_{k/k-1}^x C_k^T H_k^{-1}\left(I - E_k E_k^*\right) + \Gamma_k E_k^* \quad (42)$$

**Proof:**
According to equation (16) and after (42), we can deduce that
$$\hat{x}_{k/k} = M_k^{22} \hat{x}_{k/k-1} + M_k^{21}\left(y_k - C_k \hat{x}_{k/k-1}\right) \quad (43)$$

From (24), we know that $M_k^{22} = I_n$ then we have:
$$\hat{x}_{k/k} = \hat{x}_{k/k-1} + M_k^{21}\left(y_k - C_k \hat{x}_{k/k-1}\right) \quad (44)$$

Using (44) the state estimation error, given by
$$\tilde{x}_k = x_k - \hat{x}_k \quad (45)$$
$$\tilde{x}_k = \left(I - M_k^{21} C_k\right)\bar{\tilde{x}}_{k/k-1} - M_k^{11} F_k^y f_k - \left(M_k^{11} C_k G_{k-1} - G_{k-1}\right) d_{k-1} - M_k^{21} v_k$$

Considering (28) and (45), we determine $P_k^x$ as follows:
$$P_k^x = \left(I - M_k^{21} C_k\right) \bar{P}_{k-1}^x \left(I - M_k^{21} C_k\right)^T + M_k^{21} R_k \left(M_k^{21}\right)^T$$
$$= M_k^{21} H_k \left(M_k^{21}\right)^T - 2 \bar{P}_{k/k-1}^x H_k^T \left(M_k^{21}\right)^T + \bar{P}_{k/k-1}^x \quad (46)$$

So, the optimization problem can be solved using Lagrange multipliers
$$\text{trace}\left\{M_k^{21} C_k \left(M_k^{21}\right)^T - 2\bar{P}_{k/k-1}^x C_k^T \left(M_k^{21}\right)^T + \bar{P}_{k/k-1}^x\right\} - 2\text{trace}\left\{\left[\left(M_k^{21}\right)^T E_k - \Gamma_k\right] \Lambda_k^T\right\} \quad (47)$$

where $\Lambda_k$ is the matrix of Lagrange multipliers.

Setting the derivate of (47) with respect to $M_k^{21}$ we obtain:
$$H_k \left(M_k^{21}\right)^{*T} - C_k \bar{P}_{k/k-1} - E_k \Lambda_k^T = 0 \quad (48)$$

Equation (28) and (48) form the linear systems of equation

$$\begin{bmatrix} H_k & -E_k \\ E_k^T & 0 \end{bmatrix} \begin{bmatrix} \left(M_k^{21}\right)^T \\ \Lambda_k^T \end{bmatrix} = \begin{bmatrix} C_k \bar{P}_{k/k-1}^x \\ \Gamma_k^T \end{bmatrix} \quad (49)$$

So, if $\left(E_k^T H_k^{-1} E_k\right)$ is non singular, (49) will have unique solution.

### B. The filter time update

For the time update, we extract from (11) the equation that depend on $x_{k+1}$ and substitute $x_k$ and $f_k$ for their LS estimates $\hat{x}_{k/k}$ and $\hat{f}_{k/k}$ obtained during the measurement update. This yield,
$$A_k \hat{x}_{k/k} + F_k^x f_k + B_k u_k = x_{k+1} - \left(A_k \tilde{x}_{k/k} + F_k^x \tilde{f}_k + w_k\right) \quad (50)$$

The corresponding LS problem is given by
$$\min_{x_{k+1}} \left\| x_{k+1} - A_k \hat{x}_{k/k} - F_k^x f_k - B_k u_k \right\|_{\bar{W}_{3,k}}^2 \quad (51)$$

where $\bar{W}_{3,k}$ denotes the weighting matrix which we choose
$$\bar{W}_{3,k} = \varepsilon\left[\left(A_k \tilde{x}_{k/k} + F_k^x \tilde{f}_k + w_k\right)\left(A_k \tilde{x}_{k/k} + F_k^x \tilde{f}_k + w_k\right)^T\right]^{-1} \quad (52)$$

From equation (51), we have
$$\hat{x}_{k+1/k} = A_k \hat{x}_{k/k} + F_k^x \hat{f}_k + B_k u_k \quad (53)$$

From equation (32), the prior covariance $\bar{P}_{k/k-1}^x$ has the following form:
$$\bar{P}_{k/k-1}^x = \begin{bmatrix} A_{k-1} & F_{k-1}^x \end{bmatrix} \begin{bmatrix} P_{k-1}^{x*} & P_{k-1}^{xf*} \\ P_{k-1}^{fx*} & P_{k-1}^{f*} \end{bmatrix} \begin{bmatrix} A_{k-1}^T \\ \left(F_{k-1}^x\right)^t \end{bmatrix} + Q_{k-1} \quad (54)$$

Where $P_{k/k}^{xf*} = \varepsilon\left[\tilde{x}_k^* \tilde{f}_k^{*T}\right]$ is calculated by using (16)
$$P_{k/k}^{xf*} = -\left(I - \left(M_k^{21}\right)^* C_k\right) \bar{P}_{k/k-1}^x C_k^T \left(M_k^{21}\right)^{*T} + M_k^{22} R_k \left(M_k^{22}\right)^{*T} \quad (55)$$

### 4. EXTENDED FILTER

In this section, we seek to extend this filter to consider the case where $0 < \text{rang}\left(F_k^y\right) \leq p$. To solve this problem, we use the same approach developed by [10]. If we introduce (31) et (32) in (39), then we will be able to write the fault error estimation in the following form:
$$\tilde{f}_k = \left(I_n - M_k^{11} F_k^y\right) f_k - M_k^{11} C_k G_{k-1} d_{k-1} - M_k^{11}\left(C_k \bar{\tilde{x}}_{k/k-1} + v_k\right) \quad (56)$$
$$= -M_k^{11} C_k F_{k-1}^x \tilde{f}_{k-1} - M_k^{11} C_k A_{k-1} \tilde{x}_{k-1} + \left(I_n - M_k^{11} F_k^y\right) f_k$$
$$- M_k^{11} C_k G_{k-1} d_{k-1} - M_k^{11} C_k w_{k-1} - M_k^{11} v_k$$

Assuming that $\varepsilon\left[\tilde{x}_{k-1}\right] = 0$ we define the following notations:
$$\Phi_k = M_k^{11} F_k^y = I_p - \Sigma_k \ , \ E_k^f = M_k^{11} C_k F_{k-1}^x \quad (57)$$
$$E_k^d = M_k^{11} C_k G_{k-1} \text{ where } \Sigma_k = I - \left(F_k^y\right)^+ \left(F_k^y\right) \quad (58)$$

Using the same technique presented in [9] the expectation value of the $\tilde{f}_k$ is given by:

$$\varepsilon[\tilde{f}_k] = \Sigma_k f_k - E_k^f \Sigma_{k-1} f_{k-1} + E_k^f (E_k \Sigma_{k-2}) f_{k-2} + \cdots + (-1)^k E_k^f \times \cdots \times E_2^f$$
$$\times (E_1^f \Sigma_0) f_0 - E_k^d d_{k-1} + E_k^f E_{k-1}^d d_{k-2} + \cdots + (-1)^k E_k^f \times \cdots \times E_1^f E_1^d d_0 \quad (59)$$

We assume that $E_i^f \Sigma_{i-1} = 0$ and $E_i^d = 0$ for $i = 1, \ldots, k$, then we obtain:
$$\varepsilon[\tilde{f}_k] = \Sigma_k f_k \quad (60)$$

To obtain an unbiased estimation of the fault, the gain matrix $M_k^{11}$ should respect the following constraints :
$$M_k^{11} F_k^y = \Phi_k, \; M_k^{11} C_k F_{k-1}^x \Sigma_{k-1} = 0, \; M_k^{11} C_k G_{k-1} = 0 \quad (61)$$

The equation (61) can be writen as
$$M_k^{11} \bar{E}_k = \bar{T}_k \quad (62)$$

where $\bar{T} = [\Phi \; 0 \; 0], \bar{E}_k = [F_k^y \;\; C_k F_{k-1}^x \Sigma_{k-1} \;\; C_k G_{k-1}] \quad (63)$

Using (63), we can determie the gain matrix $M_k^{11}$ as follows:
$$M_k^{11} = \bar{T}_k \bar{E}_k^* \text{ where } \bar{E}_k^* = (E_k^T H_k^{-1} \bar{E}_k)^+ \bar{E}_k^T H_k^{-1} \quad (64)$$

The state estimation error is given in the following form :
$$\tilde{x}_k = (I - M_k^{21} C_k) \tilde{\bar{x}}_{k/k-1} - M_k^{21} F_k^y f_k - (M_k^{21} C_k G_k - G_{k-1}) d_{k-1} - M_k^{21} v_k$$
$$= (I - M_k^{21} C_k) A_{k-1} \tilde{\bar{x}}_{k/k-1} + (I - M_k^{21} C_k) F_{k-1}^x \tilde{f}_{k-1} - M_k^{21} F_k^y f_k - (M_k^{21} C_k G_k - G_{k-1}) d_{k-1}$$
$$+ (I - M_k^{21} C_k) w_{k-1} - M_k^{21} v_k \quad (65)$$

To obtain an unbiased estimate of the state, the gain $M_k^{21}$ should be satisfy the following constraints:
$$M_k^{21} F_k^y = 0, \; M_k^{21} C_k F_{k-1}^x \Sigma_{k-1} = F_{k-1}^x \Sigma_{k-1} \quad (66)$$
$$M_k^{21} C_k G_{k-1} = G_{k-1} \quad (67)$$

From (66)-(67) we obtain:
$$M_k^{21} \bar{E}_k = \bar{\Gamma}_k, \text{ where } \bar{\Gamma}_k = [0 \;\; F_{k-1}^x \Sigma_{k-1} \;\; G_{k-1}] \quad (68)$$

Refer to (65), the error state covariance matrix is given in following form
$$P_k^x = -(I - M_k^{21} C_k) \bar{P}_{k/k-1}^x (I - M_k^{21} C_k)^T + M_k^{21} C_k M_k^{21^T} \quad (74)$$
$$P_k^x = M_k^{21} C_k M_k^{21^T} - 2\bar{P}_{k/k-1}^x C_k^T M_k^{21^T} + \bar{P}_{k/k-1}^x \quad (75)$$

The gain matrix $M_k^{21}$ is determin by minimizing the trace of the covariance matrix $P_k^x$ such as (67).
$$(M_k^{21})^* = \bar{P}_{k/k-1}^x C_k^T H_k^{-1} (I - \bar{E}_k \bar{E}_k^*) + \bar{\Gamma}_k \bar{E}_k^* \quad (76)$$

Updating the filter is given by the equations (53) - (54)

## 5. AN ILLUSTRATIVE EXAMPLE

We consider the same numerical example used in [14]. The linearized model of a simplified longitudinal flight control systems is the following:
$$x_{k+1} = (A_k + \Delta A_k) x_k + (B_k + \Delta B_k) u_k + F_k^a f_k^a + w_k$$
$$y_k = C_k x_k + F_k^s f_k^s + v_k$$

where the state variable are pitch angle $\delta_z$, pitch rate $w_z$ and normal velocity $\eta_y$, the control input is elevator control signal. $F_k^a$ and $F_k^s$ are the matrices distribution of the actuator fault $f_k^a$ and sensor fault $f_k^s$.

The presented systems equation can be rewritten as follow:
$$x_{k+1} = A_k x_k + B_k u_k + F_k^a f_k^a + G_k d_k + w_k$$
$$y_k = C_k x_k + F_k^s f_k^s + v_k$$

Where $F_k^a$ and $F_k^s$ are the matrices injection of the faults vector in the same and measurement equations.
$$F_k^x = [F_k^a \;\; 0], F_k^y = [0 \;\; F_k^s]$$

The term $G_k d_k$ represents the parameter perturbation in matrices $A_k$ and $B_k$.
$$G_k d_k = \Delta A_k x_k + \Delta B_k u_k$$

The system parameter matrices are:
$$A_k = \begin{bmatrix} 0.9944 & 0.1203 & -0.4302 \\ 0.0017 & 0.9902 & -0.0747 \\ 0 & 0.8187 & 0 \end{bmatrix}, B_k = \begin{bmatrix} 0.4252 \\ -0.0082 \\ 0.1813 \end{bmatrix},$$

$$C_k = I_{3\times3}, x_k = [\delta_z \;\; w_z \;\; \eta_y]^T, R_k = 0.1^2 \, eye(4)$$
$$Q_k = diag\{0.1^2, \; 0.1^2, \; 0.01^2\}$$

We inject simultaneously two faults in the systems,
$$\begin{bmatrix} f_k^a \\ f_k^s \end{bmatrix} = \begin{bmatrix} 4u_s(k-20) - 4u_s(k-60) \\ -2u_s(k-30) + 2u_s(k-70) \end{bmatrix}$$

where $u_s$ is the unit-step function. The first fault $f_k^a$ occus in the actuator and the second fault $f_k^s$ occus in the sensor $\delta_z$

The unknown a disturbance is given by:
$$G_k d_k = G_k \left\{ \begin{bmatrix} \Delta a_{11} & \Delta a_{12} & \Delta a_{13} \\ \Delta a_{21} & \Delta a_{22} & \Delta a_{23} \end{bmatrix} x_k + \begin{bmatrix} \Delta b_1 \\ \Delta b_2 \end{bmatrix} u_k \right\}$$

where $\Delta a_{ij}$ and $\Delta b_{ij}$ $(i=1,2 \; ; j=1,2,3)$ are perturbations in aerodynamic and control coefficients.

The matrices injections of the fault and unknown disturbances are taken as follows:
$$G_k = \begin{bmatrix} 0 \\ 1 \\ 0 \end{bmatrix}, F_k^a = \begin{bmatrix} 0.4252 \\ -0.0082 \\ 0.1813 \end{bmatrix}, F_k^s = \begin{bmatrix} 0 \\ 0 \\ 1 \end{bmatrix},$$

In this simulation, the aerodynamic coefficients are perturbed by $\pm 50\%$ , i.e $\Delta a_{ij} = -0.5 a_{ij}$ and $\Delta b_{ij} = 0.5 b_{ij}$.

In addition, we set $u_k = 10$, $x_0 = [0 \; 0 \; 0]^T$, $P_0 = 0.1^2 \, eye(3)$

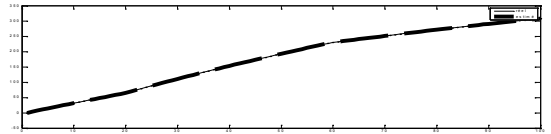

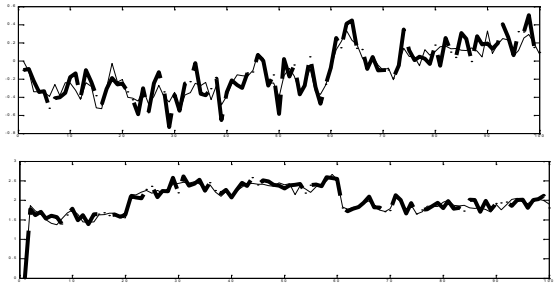

**Fig1. Actual state $x_k$ and estimated $\hat{x}_k$**

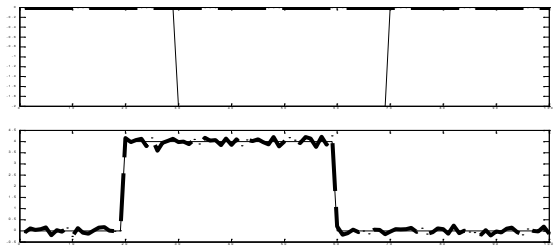

**Fig2. Actual fault $f_k$ and estimated $\hat{f}_k$**

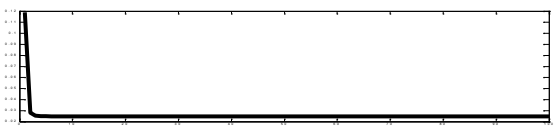

**Fig3. Trace of the covariance matrix $P_k^x$**

Figures 1 and 2 present the actual state and fault vector and theirs estimated values obtained by the proposed filter. Convergence of the trace of the state covariance matrix and fault covariance matrix are shown respectively in Fig 3.

## CONCLUSION

In this paper, the robust filter is developed to obtain an effective state and fault estimation of linear stochastic system in presence of unknown input. The advantages of this filter are especially important in the case when we do not have any prior informations about the unknown disturbances and fault. An application and the robustness of the proposed filter has been shown by an illustrative example.

## REFERENCES

[1] B. Friedland: Treatment of bias in recursive filtering. IEEE Transactions Control, vol. 14, pp.359-367, 1969.

[2] P. K. Kitanidis: Unbaised minimum variance linear state estimation. Automatica, vol.23, no. 6, pp. 775-778 1987.

[3] M. Darouach, M. Zasadzinki, and M. Boutayeb: Extension of minimum variance estimation for systems with unknown inputs. Automatica, vol.39, no.5 pp.867-876, 2003.

[4] C.S. Hsieh: Robust twi-stage Kalman filters for systems with unknown inputs. IEEE Transaction on Automatica Control, vol. 45, no. 12, pp. 2374-23778 2000.

[5] C.S. Hsieh: Optimal minimum–variance filtering for systems with unknown inputs. In proccedings of the 6th World Congress on Intelligent Control and Automatica (WCICA'06), vol.1 Dalian, Ghina, pp.1870-1874, 2006.

[6] S. Gillijns and B. Moor: Unbaised minimum-variance input and state estimation for linear siscret-time systems with direct feedthrough. Automatica, vol.43, no.5, pp.934-937, 2007.

[7] S. Gillijns and B. Moor: Unbaised minimum-variance input and state estimation for linear siscret-time systems. Automatica,vol.43, no.1, pp.111-116 2007.

[8] Y. Cheng, H. Ye, Y. Wang, and D. Zhou: Unbaised minimum-variance state estimation for linear discret-time systems with unknown input. Automatica, vol.45,no. 2, pp 485-491 2009.

[9] C.S. Hsieh: Extension of unbiased minimum-variance input and state estimation for systems with unknown input. Automatica, vol.45, no pp. 2149-2153, (2009).

[10] B. Talel and B.H. Fayçal : Recursive Least–Squares Estimation for the joint input-state estimation of linear discrete time systems with unknown input , 8th International Multi- Conference on Systems , Signals & devices, Hammamet Tunisia,2011;

[11] Nikoukhah, R. (1994). Innovation generation in the presence of unknown inputs: Application to robust failure detection, Automatica 30(12): 1851–1867.

[12] Keller, J. (1998). Fault isolation filter design for linear Stochastic systems with unknown inputs, 37th IEEEConference on Decision and Control, Tampa, FL, USA, pp. 598–603.

[13] Chen, J. and Patton, R. (1996). Optimal filtering and robust fault diagnosis of stochastic system with unknown disturbances, IEE Proceedings: Control Theory Application 143(1): 31–36.

[14] Chen, J. and Patton, R. (1999). Robust Model-based Fault Diagnosis for Dynamic Systems, Kluwer Academic Publishers, Norwell, MA.

[15] Ben Hmida, F., Khémiri, K., Ragot, J. and Gossa, M. (2010). Unbiased minimum-variance filter for state and fault estimation of linear time-varying systems with unknown disturbances, Mathematical Problems in Engineering, Vol. 2010, Article ID 343586, 17 pages.

[16] Jamouli, H., Keller, J. and Sauter, D. (2003). Fault isolation filter with unknown inputs in stochastic systems, Proceedings of Safeprocess, Washington, DC, USA, pp. 531– 536.

[17] K.hhméri, F;Ben hmida " Novel optimal recursive filter for state and fault estimation of linear stochastic systems with unknown disturbances. Int. J. Appl. Math. Comput. Sci., 2011, Vol. 21, No. 4, 629–637.

[18] Golub and Van Loan , "G. Golub & C. van Loan Matrix computations. Third edition. London: The Johns Hopkins University Press , 1996